\documentclass[a4paper,12pt]{article}

\usepackage[english]{babel} 
\usepackage{times}

\def \No{No.}
\usepackage{latexsym} 
\usepackage{amsmath}
\usepackage{amssymb}
\usepackage{graphicx}
\usepackage{wrapfig}
\usepackage{sectsty}
\usepackage{indentfirst}
\allsectionsfont{\centering}
\makeatletter
\def\@seccntformat#1{\csname the#1\endcsname.\quad}
\makeatother

\renewcommand{\thefootnote}{\fnsymbol{footnote}}

\newcommand{\arctg}{\mathop{\rm arctg}\nolimits}
\newcommand{\sh}{\mathop{\rm sh}\nolimits}
\newcommand{\ch}{\mathop{\rm ch}\nolimits}
\renewcommand{\th}{\mathop{\rm th}\nolimits}
\newcommand{\ctg}{\mathop{\rm ctg}\nolimits}
\newcommand{\tg}{\mathop{\rm tg}\nolimits}
\newcommand{\grad}{\mathop{\rm grad}\nolimits}
\newcommand{\rot}{\mathop{\rm rot}\nolimits}
\newcommand{\arcsh}{\mathop{\rm arcsh}\nolimits}
\newcommand{\tot}{\mathop{\rm tot}\nolimits}

\def\s#1{\sqrt{#1}}

\newcommand{\be}{\begin{equation}}
\newcommand{\ee}{\end{equation}}
\newcommand{\ba}{\begin{eqnarray}}
\newcommand{\ea}{\end{eqnarray}}
\newcommand{\pa}{\partial}

\newcommand{\st}{\stackrel}
\newcommand{\tk}{\tilde\kappa}
\newcommand{\ep}{\epsilon}
\newcommand{\ds}{\displaystyle}

\begin{document}

\title{\LARGE\textbf{Repulsive force in the field theory of gravitation}}
\date{}
\author{\normalsize S.\,S.~Gershtein, A.\,A.~Logunov,
M.\,A.~Mestvirishvili}
               
\date{}
\maketitle
\begin{abstract}
 It is shown that the slowing down of the rate of time referencing to the inertial time leads in the field theory
of gravitation to arising of repulsive forces which remove the cosmological singularity in the evolution of a
ho\-mo\-ge\-ne\-o\-us and isotropic universe and stop the collapse of large masses.
\end{abstract}


{ It is shown that the slowing down of the rate of time referencing to the inertial time leads in the field theory
of gravitation to arising of repulsive forces which remove the cosmological singularity in the evolution of a
ho\-mo\-ge\-ne\-o\-us and isotropic universe and stop the collapse of large masses.}

\vspace * {-3mm}

Both in the Newton theory of gravitation, and in the General Theory of Relativity (GRT) the gravitational force is exclusively an attractive
one. However field notions of gravitation show that in strong
gravitational fields it is not absolutely so. But this will be discussed below.

In the Relativistic Theory of Gravitation (RTG) [1,2]  the gravitational field
is considered as a physical field $ \phi ^ {\mu\nu} $ with spins $ 2 $ and $ 0
$, developing as well as all other physical fields in the Minkowski space. It
means that the Special Relativity Theory lays in the basis of RTG, and
consequently, the relativity principle has a universal meaning. It is
valid for all physical effects, including gravitational ones. This circumstance
ensures validness of energy-momentum and angular momentum conservation laws for
all physical processes, including gravitational ones. The RTG starts with a
hypothesis, that  gravity is universal and  the conserved
energy-momentum tensor of all the substance fields, including the
gravitational one, is the source of it.

Such an approach is in accordance with the Einstein idea.  He wrote on it
still in 1913~[3]: \textit {``\ldots the gravitational field  tensor $
\vartheta _ {\mu\nu} $ is a source of the field in parallel with the tensor of
material systems $ \Theta _ {\mu\nu} $. The exclusive position of the energy of
the gravitational field in comparison with all other sorts of energy would result
in intolerable consequences.''} Just this idea by Einstein was put in the
basis of the Relativistic Theory of Gravitation set-up.  Einstein had not succeded to implement this idea in constructing the General Relativity Theory, as
the pseudotensor of the gravitational field had
appeared in the GRT instead of the energy-momentum tensor of the gravitational field. All this had occured because Einstein did not consider the gravitational field as a physical one in the Minkowski space (in the Faraday - Maxwell meaning). For this reason the GRT does
not contain the Minkowski space metric in its equations.

The approach to gravitation accepted in the RTG leads to the \textbf{geo\-met\-ri\-za\-ti\-on}: there is an
effective Riemannian space, \textbf{but only with trivial topology}. This leads to the following picture: the
motion of a test body in the Minkowski space under the action of a gravitational field is equivalent to a motion of
this body in the effective Riemannian space created by this gravitational field. The forces of gravitation are
physical ones, and therefore they can not be reduced to zero by a choice of coordinate system. Just this  allows
to separate the inertial forces from the forces of gravitation in this theory. 
There is an effective Riemannian
space in the field approach to gravitation, but with a trivial topology only. For this reason the field notions
can not lead us to the GRT, where the topology is non-trivial in general case.

The notions described above result in the following complete set of equa\-ti\-ons [1,2]:
\be
\Bigl
(R^{\mu\nu}-\frac{\, 1 \,}{2} g^{\mu\nu} R\Bigr) + \frac{m^2}{2} \Bigl [g^{\mu\nu} + \Bigl(g^{\mu\alpha}
g^{\nu\beta} -\frac{\, 1 \,}{2} g^{\mu\nu} g^{\alpha\beta}\Bigr)\gamma_{\alpha\beta}\Bigr] =8\pi GT^{\mu\nu}\, ,
\label{eq1}
\ee
\be
D_\nu\tilde{g}^{\nu\mu}=0. \label{eq2}
\ee
Here $D_\nu $ is the covariant derivative in
Minkowski space;
$\gamma_{\alpha\beta}$
is the Minkowski space metric tensor; $g_{\alpha\beta}$ is the effective
Riemannian space metric tensor; $m=m_gc/\hbar $, $m_g $ is the graviton mass; $\tilde{g}^{\nu\mu} = \sqrt {-g} \,
g ^ {\nu\mu} $ is the Riemannian space metric tensor $g ^ {\nu\mu} $ density.

The effective metric of Riemannian space $g ^ {\mu\nu} $ is bound to the
gravitational field $ \phi ^ {\mu\nu} $ by the following relation
\[
\tilde{g}^{\mu\nu} = \tilde {\gamma}^{\mu\nu} + \tilde {\phi} ^ {\mu\nu},
\]
where
\[
\tilde {\gamma} ^ {\mu\nu} = \sqrt {-\gamma}\, \gamma ^ {\mu\nu}, \quad \tilde
{\phi} ^ {\mu\nu} = \sqrt {-\gamma}\, \phi ^ {\mu\nu}.
\]

The set of equations (\ref{eq1})--(\ref{eq2}) is covariant concerning arbitrary trans\-for\-ma\-ti\-on of
coordinates and is form-invariant concerning Lorentz trans\-for\-ma\-ti\-ons. This set can be derived
directly from the least action principle for the Lagrangian density

\[
L=L_g (\gamma _ {\mu\nu}, \tilde {g} ^ {\mu\nu})
+L_M (\tilde {g} ^ {\mu\nu}, \phi_A),
\]
where
\[
L_g =\frac {1} {16\pi} \tilde {g} ^ {\mu\nu} (G _ {\mu\nu} ^ \lambda G ^\sigma _ {\lambda\sigma} -G _ {\mu\sigma}
^ \lambda G ^\sigma _ {\nu\lambda}) -\frac {m^2} {16\pi} \Bigl (\frac {\ 1 \ } {2} \gamma _ {\mu\nu} \tilde {g} ^
{\mu\nu} -\sqrt {-g} -\sqrt {-\gamma} \Bigr),
\]
\[
G _ {\mu\nu} ^ \lambda = \frac {\, 1 } {2} g ^ {\lambda\sigma} ( D_\mu g _ {\sigma\nu} +D_\nu g _ {\sigma\mu}
-D_\sigma g _ {\mu\nu}) .
\]

To guarantee that timelike and isotropic line intervals of the effective Riemannian space did
not cross the lightcone of initial Minkowski space, the following causality condition should be satisfied:
\be
\gamma _ {\mu\nu} U ^\mu U ^\nu =0,\quad
g _ {\mu\nu} U ^\mu U ^\nu\le 0,
\label {eq3}
\ee
Thus, the motion of test bodies under the action of gravitational field
always happens  both \textbf{inside} the Riemannian cone, and also inside the  cone of  Minkowski
space.

The graviton rest-mass  appears in this theory with necessity, as only with
its introduction it is possible to consider the gravitational field as a physical
field in  Minkowski space, taking as its source the total conserved
energy-momentum tensor of all substance. But the presence of  graviton rest-mass completely changes both the process of collapse and evolution of the
Universe.

When A.~Einstein in 1912 has connected the gravitational field with
the Riemannian space metric tensor, it has appeared, that such a field gives rise to slowing down of the rate of time for a physical process. It is possible to illustrate this
slowing down, in particular, by the example of Schwarzschild solution,
comparing a rate of time in the presence of a gravitational field with the
rate of time for the removed observer. However only the Riemannian space metric tensor is present in
the GRT in general, and so, there are no any traces of the Minkowski space inertial time in the Hilbert-Einstein equations.
For this reason the universal property of gravitational
field to slow down the rate of time in comparison with the
inertial time could not be developed further in the GRT.

The origin of effective Riemannian space in the field theory of gravitation at
preserving the Minkowski space as a basic space adds the special meaning to the property of a
gravitational field to slow down the rate of time. Just only in
this case  it is possible to speak in full about slowing down of the
rate of time, realizing comparison of the rate of time in a gravitational
field with the rate of time
$T$ in the inertial system of coordinates of  Minkowski space at lack of
gravitation. All this is just implemented in RTG, as the complete set of
its equations is entered by the Minkowski space metric tensor $\gamma_{\mu\nu}$. But this general property of the gravitational field --- to slow down the rate  of time --- leads in the field theory to an important
deduction [4]: \textbf {the slowing down of the rate of time of a physical
process in a strong gravitational field in comparison with the rate of inertial
time $ \boldmath T $ creates, due to the graviton rest-mass, effective field
forces of the gravitational nature. These effective forces in the case of gravity appear
as repulsive ones}.

To show, that the modification of a rate of time results in arising of a
force, let us look at the Newton equations:
\[
m\frac {d^2x} {dT^2} =F.
\]
If we pass formally in this equation  from inertial time $T$ to time $\tau$ by a rule \be d\tau =U(T)dT,
\label{eq4} \ee then it is easy to get 
\be m\frac {d^2x} {d\tau^2} = \frac {1} {U^2} \Bigl\{ F-\frac {dx} {dT} \frac
{d} {dT} \ln U\Bigr\}.  \label {eq5} 
\ee 
From here it is evident that a modification of the  rate of time,
determined by function $U$, results in appearance of an effective force. But here all this  has only a formal
character, as in this case there is no physical effect that would change the rate of time. But this formal example
demonstrates, that if there is a real  process of slowing down of the rate of time, it inevitably creates
effective field forces, and therefore they should be taken into account in the theory as something completely new
and surprising. The physical gravitational field changes both rate of time, and parameters of spatial components,
in comparison with the same components in an inertial system of a Minkowski space at lack of gravitation.

In the present paper we will explicitly consider both the collapse and the evolution of the homogeneous and
isotropic Universe as examples where the effective field repulsive forces originating due to slowing down of the
rate of time under the effect of a gravitational field are exhibited. Let us consider a static spherically
symmetric field \be ds^2 =U (r) dT^2 -V (r) dr^2-W^2 (r) ( d\theta^2 +\sin^2\theta \, d\phi^2), \label {eq6} \ee
\be d\sigma^2 =dT^2 -dr^2-r^2 ( d\theta^2 +\sin^2\theta\, d\phi^2). \label{eq7} \ee Here function $U$ determines
slowing down of the rate of time in comparison with the inertial time~$T$. The strong slowing down of the rate
of time occurs  when this function is small enough in comparison to unity. When the graviton has no mass the set
of equations (\ref {eq1}), (\ref {eq2}) for the problem (\ref {eq6}) has the Schwarz\-schild solution \be \label
{eq8} U = \frac {r-GM} {r+GM}, \quad V = \frac {r+GM} {r-GM}, \quad W = (r+GM) . \ee From here it is evident that
the strong slowing down of the rate of time in comparison with the inertial time $T $ takes place in the region
where $W $ is close to $2GM $. At presence of the graviton rest-mass  the set of equations (\ref {eq1}) and (\ref
{eq2}) reduces [see Appendix $A $: $ (A.61), \, (A.62) $] in the region where \be \label {eq9} W-W_g\ll\frac {\, 1
\,} {2} W_g\Bigl (\frac {m_g c} {\hbar} \cdot\frac {W_g} {2} \Bigr) ^2 \ee to the following formulas: \be \label
{eq10} U = \alpha\frac {W_g} {W}, \quad V =\frac {\, 1 \,} {2} \frac {W} {W-W_g}, \quad \frac {dr} {dW} =1 \,, \ee
here \be \label {eq11} W_g =\frac {2GM} {c^2} \; \mbox {is the Schwarzschild radius}, \; \alpha =\Bigl (\frac {m_g
c} {\hbar} \cdot\frac {W_g} {2} \Bigr) ^2 . \ee Comparing (\ref {eq8}) with (\ref {eq10}), we see, that the
graviton mass  $m_g $ does not allow function $U $ to be  zero. \textbf {For any body the graviton rest-mass puts
its own limit onto slowing down of the rate of time}. This limit is determined by a linear function of the
Schwarzschild radius i.e. of the body mass  \[ \frac {\, 1 \,} {2} 
\Bigl(\frac {m_gc} {\hbar} \Bigr) W_g .
\]
There is no such a limit in the GRT. Such property of the gravitational field leads in RTG to cardinal
modifications both in a test body motion in the gravitational field, given by expressions (\ref {eq10}), and in
the evolution of the homogeneous and isotropic Universe.

The motion of a test body occurs along a geodesic line of the Riemannian space \be \label{eq12} \frac{dv^\mu}{ds} + \varGamma_{\alpha\beta}^\mu
\frac{dx^\alpha}{ds} \cdot\frac{dx^\beta}{ds} =0 \,, \ee here $v
^\mu=dx ^\mu /ds $ is the four-vector of  velocity $v ^\mu $ obeying the requirement \be \label {eq13} g _ {\mu\nu}
v ^\mu v ^\nu =1 . \ee

Let us consider the radial motion \be \label {eq14} v ^\theta =v ^\phi =0, \quad v^1 =dr/ds . \ee Taking into
consideration that the Christoffel symbol $ \varGamma _ {01} ^0 $ is equal to the following \be \label {eq15}
\varGamma _ {01} ^0 =\frac {1} {2U} \frac {dU} {dr} , \ee from  Eq.~(12) we get \be \label {eq16} \frac
{dv^0} {ds} + \frac {1} {U} \frac {dU} {dr} v^0 v^1 =0 . \ee Solving  Eq.~(16), we obtain \be \label {eq17}
\frac {d} {dr} \ln (v^0 U) =0 . \ee From here we have \be \label {eq18} v_0 =\frac {dx^0} {ds} = \frac {U_0} {U}
, \ee where $U_0$ is an integration constant. If we take the velocity of a falling test body equal to zero at
infinity, we shall receive $U_0=1 $. From  relation (13) it is discovered \be \label {eq19} \frac {dr} {ds} =
-\sqrt {\frac {1-U} {UV}} . \ee Substituting (10) into this expression, we obtain \be \label {eq20} \frac {dW}
{ds} = -\Bigl (\frac {\hbar} {m_g c} \Bigr) \frac {2} {W_g} \sqrt {2\frac {W} {W_g} \Bigl (1-\frac {W_g} {W}
\Bigr)} . \ee From here it is evident that there is a turning point. Differentiating (20) by $s $, we get \be
\label {eq21} \frac {d^2 W} {ds^2} =4\Bigl (\frac {\hbar} {m_g c} \Bigr) ^2\frac {1} {W_g^3} . \ee We see that
the acceleration is positive at the turning point, i.e. a repulsion takes place and it is significant. Integrating
(20) we obtain \be \label {eq22} W=W_g+2\Bigl (\frac {\hbar} {m_g c} \Bigr) ^2\frac {(s-s_0) ^2} {W_g^3} . \ee
From expression (22) it is clear that the test body can not cross the Schwarz\-schild sphere.

As the singularity in Eq.~(10), which has arisen outside of matter, is impossible to remove by a choice of the
frame, this means, that it should not exist, as otherwise it is impossible to sew together a solution inside
matter with an exterior solution. Therefore a body can not have radius less than  the Schwarzschild radius. So
there is a restriction on the field magnitude.

This conclusion on the absence of the Schwarzschild singularity is in accordance with the inference by Einstein
[5]: \textit {``The Schwarzschild singularity is absent, as the substance cannot be concentrated arbitrarily;
otherwise par\-tic\-les providing accumulations, will reach the light velocity''.} Though in our case the reason for
the lack of singularity is another, but a general conclusion is the same. So, a mechanism of self-restriction is
included into the field theory  which eliminates the possibility of ``black holes'' formation.

Another example demonstrating occurrence of the new effective field forces due to a slowing down of the rate of
time is the development of a homogeneous and isotropic Universe. In this case we get the flat Universe solution
only due to  Eq.~(\ref{eq2})  [see Appendix $B$: $(B.13), \, (B.16) $], where the three-dimensional geometry is
Euclidean, i.e. \ba  ds^2 =d\tau^2 -\beta^4a^2 (\tau) (dr^2+r^2d\theta^2+r^2\sin^2\theta \, d\phi^2) \,,
\nonumber \\ [-2mm] \label {eq23} \\ [-2mm]  d\sigma^2 =\frac {1} {a^6} d\tau^2-dr^2-r^2d\theta^2
-r^2\sin^2\theta \, d\phi^2 \,, \nonumber \ea \be d\tau =a^3dT. \label {eq24} \ee As the set of equations (\ref
{eq1}) and (\ref {eq2}) is complete, this solution is unique. Giving (\ref {eq23}) Eqs.~(\ref {eq1}) can be
reduced to the following set of equations for the scale factor $a (\tau) $ [see Appendix $B $: $ (B.19), \, (B.20)
$]: \be \frac {\, 1 \,} {a} \frac {d^2 a} {d\tau^2} = -\frac {4\pi G} {3} \left (\rho + \frac {3p} {c^2} \right)
-\frac {\, 1 \,} {6} (mc) ^2\left (1-\frac {1} {a^6} \right) \,, \label {eq25} \ee \be \left ( \frac {\, 1 \,} {a}
\frac {da} {d\tau} \right) ^2 = \frac {8\pi G} {3} \rho (\tau) -\frac {1} {12} (mc) ^2 \left ( 2-\frac {3}
{a^2\beta^4} + \frac {1} {a^6} \right) . \label {eq26} \ee The scale factor $a (\tau) $ in (\ref {eq24})
determines the slowing down of the rate of time in comparison with the inertial time $T $ at lack of gravitation.
But just the same factor in r.h.s. of Eq.~(\ref{eq26}) at a strong slowing down of the rate of time stops the
process of collapse of the Universe. When $a $ becomes small enough, the term in r.h.s. of Eq.~(\ref {eq26}) \[
(mc) ^2/12a^6
\]
becomes large enough, despite of  little $m $, and  r.h.s. of Eq.~(\ref
{eq26}) becomes zero, and the collapse is stopped.

The minimum value of $a $ is equal to \be a _ {\min} = \Bigl [\Bigl (\frac {m_g c^2} {\hbar} \Bigr) ^2 \frac {1}
{32\pi G \rho _ {\max}} \Bigr] ^ {1/6} \! \!, \label {eq27} \ee thus in RTG the cosmological singularity, which
takes place in GRT, is absent. \textbf{Therefore, that there was no ``Big Bang''}, instead there was a state with a
large density $ \rho _ {\max} $ and a high temperature at every point of the Universe. On the other hand, by
virtue of Eq.~(\ref {eq25}), repulsive forces originating due to the slowing down of a rate of time  ensure
the accelerated expansion of the Universe from the point of stopping. The acceleration within the radiation phase at a
point of stopping of the contraction is given as follows 
\be 
\frac{\, 1 \,} {a} \ds\frac {d^2 a} {d\tau^2} \bigg | _
{\tau = \, 0} = \frac {8\pi G} {3} \rho _ {\max} . 
\label{eq28} 
\ee 
Just this acceleration was the ``impulse'' to
begin the expansion of Universe. The maximum value of the scale factor $a $ is equal to the following 
\[ a_{\max}=\beta .
\]
The magnitude $\beta$ is determined by an integral of motion.

The graviton rest-mass by a unique fashion has entered into Eqs.~(\ref{eq1})  metric tensor
$ \gamma _ {\alpha\beta} $ of the Minkowski space.
Due to the graviton rest-mass it is possible to determine the magnitude of slowing down of the rate of time for a
physical process
in the gravitational field in comparison to the inertial time. Just this magnitude 
has determined the  repulsive force. Due to the graviton  mass the slowing down of
the rate of time has exhibited itself as repulsive forces.

Thus, the field notions on the gravitational field evolving in the Min\-kow\-ski
space, have allowed \textbf{to unclose a fundamental property of the gra\-vi\-ta\-ti\-on\-al
field: to create an effective repulsive force, due to slowing down of the rate
of time of a physical process in comparison with the inertial time}.

In GRT such forces are absent. There is an interesting pattern: the gravitational field in RTG, exhibiting
itself through attractive forces, agg\-lo\-me\-ra\-ting matter, then enters a phase, when under effect of this field
there is a strong slowing down of the rate of time in comparison to inertial time $T $, that, due to the  graviton
rest-mass, leads inevitably to effective field repulsive forces, which stop the process of collapse
created by the action of attractive forces. We see, that in the field theory the
mechanism of self-restriction is included in the gravitational field itself. Just this mechanism realizes a stopping of the collapse of massive bodies at the
final stage of development and eliminates the cosmological singularity, ensuring a cyclical development of the
Universe.

The graviton mass  which is included in Eq.~(\ref{eq1}) can be evaluated from observational data on measuring value
 $\Omega_{\mbox{tot}}$, which is defined as the ratio of a total modern density of matter $\rho_{\mbox {tot}}$ to the critical density 
$\rho_c$ 
\be \Omega_{\mbox {tot}} = \frac{\rho_{\mbox{tot}}}
{\rho_c}, \quad \rho_c =\frac {3H^2}{8\pi G} \,, \label{eq29} \ee 
here $H $ is  the Hubble constant. From
Eqs.~(\ref{eq25})--(\ref {eq26}) it follows \be \Omega_{\mbox {tot}} =1 +
\frac{\, 1 \,}{6} \Bigl(\frac{m_gc^2}{\hbar H} \Bigr)^2 . \label {eq30} \ee The mass of graviton $m_g$ enters into $\Omega_{\mbox {tot}}$
with a large factor, determined by  value $c^2/\hbar H$.

Thus, according to RTG, as follows from Eq.~(\ref{eq30}), the modern density of matter should exceed the critical
density. So, still in 1984 in paper [6] it was scored: \textit{``This theory gives an exclusively strong
prediction
--- it results  in a strongly definite evolution of the Universe. According to it the Universe is not closed,
it is ``flat'' by virtue of the Eqs.~(4.29)} (\textrm{this means  Eq.~(\ref {eq2}) of the present article}. ---
\textsl{Authors}). And further, the theory \textit{``with necessity requires existence  ``of the hidden mass'' in the Universe, as some form of matter. So, there should be ``a hidden mass''  in the Universe, in order to
the total density of matter to be equal to the critical value $\rho_0$.''} The observational data have confirmed
this deduction recent years. With introduction of the graviton rest-mass this corollary of the theory has
strengthened and has led to formula (\ref{eq30}). Comparing (\ref{eq30}) with modern observational data on
measuring $\Omega_{\mbox {tot}}$, it is possible to find with probability 95\% the following upper bound for the
graviton mass 
\be
m_g< 3 {,} 6\cdot 10 ^ {-66} ~ \mbox {[g]} . \label {eq31} 
\ee

\vspace*{5mm} The authors express their gratitude to V.I.~Denisov, V.A.~Petrov, N.E.~Tyurin and Yu.V.~Chugreev for valuable discussions. \\

\appendix
\allsectionsfont{\raggedright}
\makeatletter
\def\@seccntformat#1{іриложение \csname the#1\endcsname.\hfill\strut}
\makeatother

\makeatletter
\renewcommand\section{%
   \if@noskipsec \leavevmode \fi
   \par
   \addvspace{4ex}%
   \@afterindentfalse
   \secdef\@xxpart\@xxspart}

\def\@xxpart[#1]#2{%
    \refstepcounter{section}%
    \addcontentsline{toc}{section}{\thesection\hspace{0.9em}#1}%
    {\parindent \z@ \raggedright
     \interlinepenalty \@M
     \normalfont
     \ifnum \c@secnumdepth >\m@ne
       \Large\bfseries іриложение\nobreakspace\thesection 
       \par\nobreak
     \fi
     \bigskip
     \centering
     \Large \bfseries #2%
     \markboth{}{}\par}%
    \nobreak
    \vskip 3ex
    \@afterheading}
\def\@xxspart#1{%
    {\parindent \z@ \raggedright
     \interlinepenalty \@M
     \normalfont
     \large \bfseries #1\par}%
     \nobreak
     \vskip 3ex
     \@afterheading}
\makeatother
\appendix

\allsectionsfont{\raggedright}

\makeatletter
\def\@seccntformat#1{іриложение \csname the#1\endcsname.\hfill\strut}
\makeatother

\makeatletter
\renewcommand\section{%
   \if@noskipsec \leavevmode \fi
   \par
   \addvspace{4ex}%
   \@afterindentfalse
   \secdef\@xxpart\@xxspart}

\def\@xxpart[#1]#2{%
    \refstepcounter{section}%
    \addcontentsline{toc}{section}{\thesection\hspace{0.9em}#1}%
    {\parindent \z@ \raggedright
     \interlinepenalty \@M
     \normalfont
     \ifnum \c@secnumdepth >\m@ne
       \Large\bfseries Appendix\nobreakspace\thesection 
       \par\nobreak
     \fi
     \bigskip
     \centering
     \Large \bfseries #2%
     \markboth{}{}\par}%
    \nobreak
    \vskip 3ex
    \@afterheading}
\def\@xxspart#1{%
    {\parindent \z@ \raggedright
     \interlinepenalty \@M
     \normalfont
     \large \bfseries #1\par}%
     \nobreak
     \vskip 3ex
     \@afterheading}
\makeatother

\section {Static spherically symmetric gravitational field in RTG}

The interval in  Minkowski space given in spatial polar coordinates looks like
$$
d\sigma^2 = (dx^0) ^2
- (dr) ^2-r^2
( d\theta^2 +\sin^2\theta \, d\phi^2) \,,
\eqno(A.1)
$$
here $x^0=cT $. The interval in effective Riemannian space for a static spherically symmetric field is written in
the following form
$$
ds^2 =U (r) (dx^0) ^2-V (r) dr^2-W^2 (r)
( d\theta^2 +\sin^2\theta \, d\phi^2) .
\eqno(A.2)
$$
The RTG equations (\ref {eq1})--(\ref {eq2}) can be taken as follows
$$
R ^\mu_\nu -\frac {\, 1 \,} {2} \delta ^\mu_\nu R +\frac {m^2} {2} \Bigl (\delta ^\mu_\nu +g ^ {\mu\alpha} \gamma
_ {\alpha\nu} -\frac {\, 1 \,} {2} \delta ^\mu_\nu g ^ {\alpha\beta} \gamma _ {\alpha\beta} \Bigr) = \varkappa T
^\mu_\nu \,, \eqno(A.3)
$$
$$
D_\mu\tilde g ^ {\mu\nu} =0 .
\eqno(A.4)
$$
In the detailed form Eq.~$(A.4)$ looks like
$$
\pa_\mu\tilde g ^ {\mu\nu} + \gamma _ {\lambda\sigma} ^ \nu\tilde g ^ {\lambda\sigma} =0 . \eqno(A.5) $$ Here $R
^\mu_\nu $ is the Ricci tensor; $R $ is the scalar curvature; $T ^\mu_\nu $ is the energy-momentum tensor of a
source; $ \gamma _ {\lambda\sigma} ^ \nu $ are Christoffel symbols of the Minkowski space; $D_\mu $ is the
covariant derivative in  Minkowski space; $G $ is the gravitational constant; $m=m_gc/\hbar $, $m_g $ is the mass
of  graviton; $ \varkappa =8\pi G/c^2 $, $ \tilde g ^ {\mu\nu} = \sqrt {-g} \, g ^ {\mu\nu} $ is the density of
tensor $g ^ {\mu\nu} $.

For a spherically symmetric static source the components of tensor $T ^\mu_\nu $ are as follows
$$
T_0^0 =\rho(r), \quad T_1^1=T_2^2=T_3^3 =-\frac {p (r)} {c^2} \,, \eqno(A.6) \vspace*{-2mm}
$$
here $ \rho $ is the mass density; $p $ is the isotropic pressure.

To determine metric coefficients $U, V $ and $W $ one can exploit Eqs.~$(A.3)$ for values of indicies $\mu =0,\, \nu=0; \, \mu=1, \nu=1 $: \vspace*{-1mm}
$$
\frac {1} {W^2} -\frac {1} {VW^2} \Bigl ( \frac {dW} {dr} \Bigr) ^2 -\frac {2} {VW} \frac {d^2W} {dr^2} -\frac {1}
{W} \frac {dW} {dr} \frac {d} {dr} \Bigl (\frac {1} {V} \Bigr) +
$$
\vspace*{-2mm}
$$
\eqno(A.7)
$$
\vspace*{-7mm}
$$
+ \frac {\, 1 \,} {2} m^2\Bigl [1 +\frac {\, 1 \,} {2} \Bigl (\frac {1} {U} -\frac {1} {V} \Bigr) -\frac {r^2}
{W^2} \Bigr] = \varkappa \rho \,,\nonumber
$$
\vspace*{3mm}
$$
\frac {1} {W^2} -\frac {1} {VW^2} \Bigl (\frac {dW} {dr} \Bigr) ^2 -\frac {1} {UVW} \frac {dW} {dr} \frac {dU}
{dr} +
$$
\vspace*{-7mm}
$$
\eqno(A.8)
$$
\vspace*{-7mm}
$$
+ \frac {\, 1 \,} {2} m^2\Bigl [1-\frac {\, 1 \,} {2} \Bigl (\frac {1} {U} -\frac {1} {V} \Bigr) -\frac {r^2}
{W^2} \Bigr] = -\varkappa \frac {p} {c^2} .
$$
Eq.~$(A.5)$ becomes
$$
\frac {d} {dr} \Bigl (\sqrt {U/V} \, W^2\Bigr) =2r\sqrt {UV} . \eqno(A.9)
$$
Taking into account the identity \[ \frac {dr} {dW} \frac {1} {W^2} \frac {d} {dr} \Bigl [\frac {W} {V} \Bigl (\frac
{dW} {dr} \Bigr) ^2\Bigr] = \frac {1} {VW^2} \Bigl (\frac {dW} {dr} \Bigr) ^2 + \frac {2} {VW} \frac {d^2W} {dr^2}
+ \frac {1} {W} \frac {dW} {dr} \frac {d} {dr} \Bigl (\frac {1} {V} \Bigr) \,,
\]
and passing from  derivatives over $r$ to derivatives over $W $,
Eqs.~$(A.7)$, $(A.8)$ and $(A.9)$ become
$$
1-\frac {d} {dW} \Bigl [\frac {W} {V (dr/dW) ^2} \Bigr] + \frac {\, 1 \,} {2} m^2\Bigl [W^2-r^2 +\frac {W^2} {2}
\Bigl (\frac {1} {U} -\frac {1} {V} \Bigr) \Bigr] = \varkappa W^2\rho \,, 
\eqno(A.10)
$$
$$
1-\frac {W} {V (dr/dW) ^2} \frac {d} {dW} [\ln (UW)] + \frac {\, 1 \,} {2} m^2\Bigl [W^2-r^2-\frac {W^2} {2} \Bigl
(\frac {1} {U} -\frac {1} {V} \Bigr) \Bigr] = -\varkappa W^2\frac {p} {c^2} \,, \eqno(A.11)
$$
$$
\frac {d} {dW} \bigl [\sqrt {U/V} \, W^2\bigr] =2r\sqrt {UV} \, \frac {dr} {dW} . \eqno(A.12)
$$
Subtracting Eq.~$(A.11)$ from  Eq.~$(A.10)$ and introducing a new variable
$$
Z = \frac {UW^2} {V\dot {r} ^2}, \quad \dot {r} = \frac {dr} {dt}, \quad t = \frac {W-W_0} {W_0} \,, \eqno(A.13)
$$
we obtain
$$
\frac {dZ} {dW} -\frac {2Z} {U} \frac {dU} {dW} -2\frac {Z} {W} -\frac {m^2W^3} {2W_0^2} \Bigl (1-\frac {U} {V}
\Bigr) = -\varkappa \frac {W^3} {W_0^2} \Bigl (\rho + \frac {p} {c^2} \Bigr) U . \eqno(A.14)
$$
After addition of  Eqs.~$(A.10)$ and $(A.11)$, we discover
$$
1-\frac {\, 1 \,} {2} \frac {W_0^2} {W} \frac {1} {U} \frac {dZ} {dW} + \frac {m^2} {2} (W^2-r^2) = \frac {\, 1
\,} {2} \varkappa W^2\Bigl (\rho -\frac {p} {c^2} \Bigr) . \eqno(A.15)
$$
Let us consider Eqs.~$(A.14)$ and $(A.15)$ outside of matter in region, defined by inequalities
$$
\frac {U} {V} \ll 1, \quad \frac {\, 1 \,} {2} m^2 (W^2-r^2) \ll 1 
. \eqno(A.16)
$$
In this region Eq.~$(A.15)$ looks like
$$
U = \frac {\, 1 \,} {2} \frac {W_0^2} {W} \frac {dZ} {dW} = \frac {\, 1 \,} {2} \frac {W_0} {W} \frac {dZ} {dt} .
\eqno(A.17)
$$
Taking into consideration $(A.17)$, we  reduce  Eq.~$(A.14)$ to the following form
$$
Z\frac {d^2Z} {dW^2} -\frac {\, 1 \,} {2} \Bigl (\frac {dZ} {dW} \Bigr) ^2 + \frac {\, 1 \,} {4} m^2\frac {W^3}
{W_0^2} \frac {dZ} {dW} =0 . \eqno(A.18)
$$
After introduction, according to $(A.13)$,  variable $t$, Eq.~$(A.18)$ takes the form
$$
Z\ddot {Z} -\frac {\, 1 \,} {2} (\dot {Z}) ^2 + \alpha (1+t) ^3\dot {Z} =0 \,, \eqno(A.19)
$$
here $ \alpha =m^2W_0^2/4, \; \dot {Z} =dZ/dt $. For values of $t$, restricted by inequality
$$
0\leq t\ll 1/3 \,,
\eqno(A.20)
$$
Eq.~$(A.19)$ is simplified:
$$
Z\ddot {Z} -\frac {\, 1 \,} {2} (\dot {Z}) ^2 + \alpha\dot {Z} =0 . 
\eqno(A.21)
$$
It has the following solution:
$$
\lambda\sqrt {Z} =2\alpha\ln\Bigl (1 +\frac {\lambda\sqrt {Z}} {2\alpha} \Bigr) + \frac {\lambda^2} {2} t \,, \eqno(A.22)
$$
here $\lambda$ is an arbitrary constant.

Given $(A.13)$ and $(A.17)$ we have
$$
U = \frac {\, 1 \,} {2} \frac {W_0} {W} \dot {Z}, \quad V\dot {r} ^2 =\frac {\, 1 \,} {2} W_0W\frac {\dot Z} {Z}
. \eqno(A.23)
$$
Using Eq.~$(A.22)$, we discover
$$
\dot {Z} =2\alpha + \sqrt {Z} .
\eqno(A.24)
$$
Substituting $(A.24)$ into $(A.23)$, we obtain
$$
U = \frac {W_0} {W} \Bigl (\alpha + \frac {\lambda} {2} \sqrt {Z} \Bigr), \quad V\dot {r} ^2=W_0W \,\frac {\alpha
+ \lambda\sqrt {Z} /2} {Z} . \eqno(A.25)
$$
At $\alpha =0$ from Eq.~$(A.22)$ we have
$$
\sqrt {Z} = \frac {\, \lambda \,} {2} t . \eqno(A.26)
$$
Substituting this expression in Eq.~$(A.25)$, we discover
$$
U = \Bigl (\frac {\, \lambda \,} {2} \Bigr) ^ {\! 2} \, \frac {W-W_0} {W} . \eqno(A.27)
$$
But this expression for $U$ should precisely coincide with the Schwarzschild solution
$$
U = \frac {W-W_g} {W}, \quad W_g =\frac {2GM} {c^2} . \eqno(A.28)
$$
Comparing $ (A.27) $ and $ (A.28) $, we obtain
$$
\lambda =2, \quad W_0=W_g .
\eqno(A.29)
$$
Thus we discover:
$$
U = \frac {W_g} {W} (\alpha + \sqrt {Z}), \quad V\dot {r} ^2=W_g W\frac {\alpha + \sqrt {Z}} {Z} . \eqno(A.30)
$$
Now it is necessary  to determine how $r$ depends on $W$ by means of Eq.~$(A.12)$.

Substituting $(A.30)$ into Eq.~$(A.12)$ and passing to a variable
$$
\ell =r/W_g \,,
\eqno(A.31)
$$
we obtain
$$
\frac {d} {d\sqrt {Z}} \Bigl [(1+t) \frac {dZ} {dt} \frac {d\ell} {d\sqrt {Z}} \Bigr] =4\ell . \eqno(A.32)
$$
Taking into account $(A.24)$ and differentiating over $\sqrt {Z}$ in Eq.~$(A.32)$, we discover
$$
(1+t) (\alpha +\sqrt {Z}) \frac {d^2\ell} {(d\sqrt {Z}) ^2} + (1+t +\sqrt {Z}) \frac {d\ell} {d\sqrt {Z}} -2\ell
=0 . \eqno(A.33)
$$
As we are interested by a range of values of $t$,  determined by inequality $(A.20)$, Eq.~$(A.33)$ in this region becomes simpler and
looks like
$$
( \alpha +\sqrt {Z}) \frac {d^2\ell} {(d\sqrt {Z}) ^2} + (1 +\sqrt {Z}) \frac {d\ell} {d\sqrt {Z}} -2\ell =0 .
\eqno(A.34)
$$
The general solution of Eq.~$(A.34)$ will be as follows
$$
\ell=A\ell_1+B\ell_2 \,,
\eqno(A.35)
$$
where
\[
\ell_1 =F [-2, \, 1-\alpha, \, - (\alpha + \sqrt {Z})], \;
\ell_2 = (\alpha +\sqrt {Z}) ^ \alpha F [-2 +\alpha, \, 1 +\alpha, \, - (\alpha
+ \sqrt {Z})] . \]
Here $A$ and $B$ are arbitrary constants;
$F$ is the degenerated hypergeometric function.

Taking into account the following equalities \[ F [-2, \, 1-\alpha, \, - (\alpha + \sqrt {Z})] =1 +\frac {2 (\alpha + \sqrt
{Z})} {1-\alpha} + \frac {(\alpha + \sqrt {Z}) ^2} {(1-\alpha) (2-\alpha)} \,,
\]
\[
F [-2 +\alpha, \, 1 +\alpha, \, - (\alpha + \sqrt {Z})]=
e ^ {- (\alpha + \sqrt {Z})} F (3, \, 1 +\alpha, \, \alpha + \sqrt {Z}) \,,
\]
we have
$$
\ell =A\Bigl [1 + \frac {2 (\alpha + \sqrt {Z})} {1-\alpha} + \frac {(\alpha + \sqrt {Z}) ^2} {(1-\alpha)
(2-\alpha)} \Bigr] -B\beta (Z) e ^ {-\alpha} F (3, \, 1 +\alpha, \, \alpha + \sqrt {Z}) . \eqno(A.36)
$$
Let us consider the following expression
$$
\frac {d\ell} {d\sqrt {Z}} =A\frac {2 (2 + \sqrt {Z})} {(1-\alpha) (2-\alpha)} +
$$
\vspace*{-7mm}
$$
\eqno(A.37)
$$
\vspace*{-7mm}
$$
+B\Bigl[-\frac {\sqrt {Z} \, F (3, \, 1 +\alpha, \, \alpha + \sqrt {Z})} {\alpha + \sqrt {Z}} + \frac {dF (3, \,
1 +\alpha, \, \alpha + \sqrt {Z})} {d (\alpha + \sqrt {Z})} \Bigr] \beta (Z) e ^ {-\alpha} \,,
$$
here
\[
\beta (Z) = (\alpha + \sqrt {Z}) ^ \alpha e ^ {-\sqrt {Z}} .
\]
But since \[ \frac {dF (3, \, 1 +\alpha, \, \alpha + \sqrt {Z})} {d (\alpha + \sqrt {Z})} = \frac {3} {1 +\alpha}
F (4, \, 2 +\alpha, \, \alpha + \sqrt {Z}) \,,
\]
we have
$$
\frac {d\ell} {d\sqrt {Z}} =A\frac {2 (2 + \sqrt {Z})} {(1-\alpha) (2-\alpha)} +
$$
\vspace*{-7mm}
$$
\eqno(A.38)
$$
\vspace*{-7mm}
$$
+B\beta (Z) \Bigl [-\frac {\sqrt {Z} \, F (3, \, 1 +\alpha, \, \alpha + \sqrt {Z})} {\alpha + \sqrt {Z}} + \frac
{3F (4, \, 2 +\alpha, \, \alpha + \sqrt {Z})} {(1 +\alpha)} \Bigr] e ^ {-\alpha} \,,
$$
As
$$
\frac {dr} {dW} = \frac {d\ell} {dt} = \frac {\alpha + \sqrt {Z}} {\sqrt {Z}} \frac {d\ell} {d\sqrt {Z}} \,, \eqno(A.39)
$$
so it is necessary for us to pick $A$ and $B$ in such a way that the
singularity of function $(A.39)$ at a point $\sqrt{Z}=0$ will be cancelled, therefore let us assume
$$
\Bigl (\frac {d\ell} {d\sqrt {Z}} \Bigr) _ {\sqrt {Z} =0} =0 . \eqno(A.40)
$$
This requirement reduces to a relation between stationary values
$$
B = -A\frac {\, 4 \,} {3} \frac {(1 +\alpha) e ^\alpha} {(1-\alpha) (2-\alpha) \alpha ^\alpha F (4, \, 2 +\alpha,
\, \alpha)} . \eqno(A.41)
$$
Substituting this expression into $(A.36)$ and $(A.37)$, we discover
$$
\ell =A\Bigl [1 + \frac {2 (\alpha + \sqrt {Z})} {1-\alpha} + \frac {(\alpha + \sqrt {Z}) ^2} {(1-\alpha)
(2-\alpha)}-
$$
\vspace*{-7mm}
$$
\eqno(A.42)
$$
\vspace*{-7mm}
$$
-\frac {\, 4 \,} {3} \frac {(1 +\alpha) F (3, \, 1 +\alpha, \, \alpha + \sqrt {Z})} {(1-\alpha) (2-\alpha) \alpha
^\alpha F (4, \, 2 +\alpha, \, \alpha)} \beta (Z) \Bigr] \,,
$$
\vspace*{3mm}
$$
\frac {d\ell} {d\sqrt {Z}}= A\frac {2} {(1-\alpha) (2-\alpha)} \Bigl [2 +\sqrt {Z} +
$$
\vspace*{-7mm}
$$
\eqno(A.43)
$$
\vspace*{-5mm}
$$
+ \frac {2\sqrt {Z} (1 +\alpha) F (3, \, 1 +\alpha, \, \alpha + \sqrt {Z})} {3\alpha ^\alpha (\alpha + \sqrt {Z})
F (4, \, 2 +\alpha, \, \alpha)} \beta (Z) -\frac {2F (4, \, 2 +\alpha, \, \alpha + \sqrt {Z})} {\alpha ^\alpha F
(4, \, 2 +\alpha, \, \alpha)} \beta (Z) \Bigr] .
$$
It is possible to show that the derivative $(A.43)$ is positive.
From expressions $(A.42)$ and $(A.43)$ by  expansion
of functions $F(3,\, 1 +\alpha, \, \alpha + \sqrt {Z})$, $F(4, \, 2
+\alpha, \, \alpha + \sqrt {Z}) $ in Taylor series  in the neighbourhood of
$\sqrt{Z}=0$, and also taking into account
\vspace*{-1mm}
\[
\alpha\ll 1,
\]
\vspace*{-1mm}
we obtain
$$
\ell =A\Bigl [1 +\frac {\, 1 \,} {2} (\sqrt {Z} + \alpha) (4 +\sqrt {Z} -\alpha) -\frac {\, 2 \,} {3} \beta (Z)
(1+3\sqrt {Z} +3Z) +O (Z ^ {3/2}) \Bigr] \,, \eqno(A.44)
$$
$$
\frac {d\ell} {d\sqrt {Z}} \simeq A\Bigl [2\Bigl (1-\beta (Z) \Bigr) + \sqrt {Z} \Bigl (1-4\beta +\frac {\, 2 \,}
{3} \frac {\beta (Z)} {\alpha +\sqrt {Z}} + \frac {2\sqrt {Z}} {\alpha +\sqrt {Z}} \beta (Z) \Bigr) +O (Z) \Bigr]
. \eqno(A.45)
$$

\pagebreak We shall consider the following limiting cases.
\begin {center}
{ \bf I.}
\end{center}
$$
\sqrt{Z}\gg\alpha .
\eqno(A.46)
$$
In this case from expression $ (A.22) $ with  account for $ (A.29) $ we have
$$
\sqrt {Z} =t .
\eqno(A.47)
$$
Substituting this expression in $ (A.25) $ and taking into account $ (A.29) $,
we obtain
$$
U = \frac {W-W_g} {W}, \quad V\dot {r} ^2 =\frac {WW_g^2} {W-W_g} . \eqno(A.48)
$$
In  approximation $ (A.46) $  expressions $ (A.44) $ and $ (A.45) $ become
$$
\ell =A\Bigl [\frac {\, 1 \,} {3} + \frac {\, 2 \,} {3} \sqrt {Z} -\frac {\, 2 \,} {3} \alpha\ln\sqrt {Z} -\frac
{\, 4 \,} {3} \alpha\sqrt {Z} \ln\sqrt {Z} + \frac {\, 1 \,} {6} Z+O (Z ^ {3/2}) \Bigr] \,, \eqno(A.49)
$$
$$
\frac {d\ell} {d\sqrt {Z}} \simeq A\Bigl [\frac {\, 2 \,} {3} + \frac {\sqrt {Z}} {3} -\frac {4\alpha} {3}
\ln\sqrt {Z} -\frac {\, 2 \,} {3} \frac {\alpha} {\sqrt {Z}} +O (\alpha) \Bigr] . \eqno(A.50)
$$

According to $(A.13)$, $ (A.31) $ and $ (A.47) $ we have
$$
\dot {r} =W_g \,\frac {d\ell} {d\sqrt {Z}}\, . \eqno(A.51)
$$
But, as in $ (A.48) $  expression for $V $ should be very close to the
Schwarzschild solution, we discover
$$
A=3/2 \,.
\eqno(A.52)
$$
So, we have the Schwarzschild solution for this case
$$
U = \frac {W-W_g} {W}, \quad V =\frac {W} {W-W_g} . \eqno(A.53)
$$

Let us pass now to another limiting case, where influence of the graviton mass  is significant. \\
\newpage
\begin{center}
{\bf II.}
\end {center}
$$
\sqrt {Z} \ll\alpha .
\eqno(A.54)
$$
In this approximation from expression $ (A.22) $ with  account for $ (A.29) $ we find
$$
Z=2\alpha t .
\eqno(A.55)
$$
Substituting this expression in $ (A.25) $ and taking into account $ (A.29) $,
we obtain
$$
U = \alpha\frac {W_g} {W}, \quad V\dot {r} ^2 =\frac {\, 1 \,} {2} \frac {WW_g^2} {W-W_g} . \eqno(A.56)
$$
According to $ (A.54) $ and $ (A.55) $  expressions $ (A.56) $ are usable in
the region
$$
t\ll\frac {\alpha} {2} \; \; \mbox {or} \; \; W-W_g\ll\frac {\, 1 \,} {2} W_g\Bigl (\frac {m_gc} {\hbar} \, \frac
{W_g} {2} \Bigr) ^2 . \eqno(A.57)
$$
For a given limiting case from $ (A.44) $ and $ (A.45) $ we have
$$
\ell \simeq A\Bigl [\frac {\, 1 \,} {3} + \frac {\alpha} {3} \Bigl (\frac {\sqrt {Z}} {\alpha} \Bigr) ^2+O (Z ^
{3/2}) \Bigr] \,, \eqno(A.58)
$$
$$
\frac {d\ell} {d\sqrt {Z}} \simeq \frac {\, 2 \,} {3} A\frac {\sqrt {Z}} {\alpha} \bigl (1+O (\sqrt {Z}) \bigr) .
\eqno(A.59)
$$
Substituting $ (A.59) $ in $ (A.39) $ and taking into account $ (A.52) $, we
discover
$$
\dot {r} =W_g \; \;\mbox {or} \; \; dr/dW=1 .
\eqno(A.60)
$$
So, for a surveyed limiting case
$ (A.54) $, taking into consideration $ (A.60) $, expressions $ (A.56) $
become as follows [7]:
$$
U = \alpha\frac {W_g} {W}\,, \quad V =\frac {\, 1 \,} {2} \frac {W} {W-W_g}\,, \quad \frac {dr} {dW} =1 \,, \eqno(A.61)
$$
they are usable in domain $ (A.57) $:
$$
W-W_g\ll\frac {\, 1 \,} {2} W_g\Bigl (\frac {m_gc} {\hbar} \, \frac {W_g} {2} \Bigr) ^2 . \eqno(A.62)
$$

Expressions $ (A.61) $ essentially differ from the Schwarzschild solution
first of all because  function $U$, which determines slowing down of the rate of
time, due to the graviton rest-mass  will not convert into zero, whereas in GRT
[see formulas (\ref{eq28})] it will be zero on the Schwarzschild sphere. If in GRT the Schwarzschild singularity is eliminated by a
choice of a reference frame, in RTG  functions $ (A.61) $ lead to a
singularity, which can not be eliminated by a coordinate transformation.
For this reason this singularity outside of matter is intolerable, as
otherwise it is impossible to sew together an interior solution with an
exterior solution. Just this results in self-restriction of a magnitude of the
gravitational field.

In conclusion of this Appendix we note that in our search for solutions
$ (A.53) $ and $ (A.61) $ we have required realization of inequalities $ (A.16) $
and $ (A.20) $.
It is easy to get convinced, that the obtained solutions $(A.53)$ and $(A.61)$
obey to inequalities $(A.16)$, when  variable $t $ is restricted accordingly
to inequalities \[
1/3\gg t\gg\alpha \,;\quad t\ll\alpha .
\]

\section{Equations for the scale factor evolution}

In the homogeneous and isotropic Universe the effective
Riemannian space interval  can be given in the form of  Fridman-Robertson-Walker metric:
$$
ds^2 =U (T) dT^2 -V (T) \Bigl [\frac {dr^2} {1-kr^2} +r^2 ( d\theta^2 +\sin^2\theta \, d\phi^2) \Bigr] \,, \eqno(B.1)
$$
The Minkowski space interval is as follows
$$
d\sigma^2 =dT^2-dr^2
-r^2 (d\theta^2 +\sin^2\theta \, d\phi^2) .
\eqno(B.2)
$$
The RTG Eqs.~(\ref {eq1})--(\ref {eq2}) we write in the following form
$$
\frac {m^2} {2} \gamma _ {\mu\nu} =8\pi G\Bigl (T _ {\mu\nu} -\frac {\, 1 \,} {2} g _ {\mu\nu} T\Bigr) -R _
{\mu\nu} + \frac {m^2} {2} g _ {\mu\nu} \,, \eqno(B.3)
$$
$$ \pa_\mu\tilde {g} ^ {\mu\nu} + \gamma _ {\lambda\sigma} ^ \nu\tilde {g} ^
{\lambda\sigma} =0 . \eqno(B.4)
$$
Taking into account that
$$ \gamma _ {\lambda\sigma} ^0=0, \; \gamma _ {22} ^1 =-r, \; \gamma _ {33} ^1
=-r\sin^2\theta, $$
$$
\tilde g ^ {00} =V ^ {3/2} U ^ {-1/2} (1-kr^2) ^ {-1/2} r^2\sin\theta,
$$
$$
\tilde g ^ {11} = - V ^ {1/2} U ^ {1/2} (1-kr^2) ^ {1/2} r^2\sin\theta, 
\eqno(B.5)
$$
$$
\tilde g ^ {22} = - V ^ {1/2} U ^ {1/2} (1-kr^2) ^ {-1/2} \sin\theta,
$$
$$
\tilde g ^ {33} = - V ^ {1/2} U ^ {1/2} (1-kr^2) ^ {-1/2} (\sin\theta) ^ {-1},
$$
Equations $ (B.4) $ for $ \nu =0 $ and $ \nu =1 $ become
$$
\frac {d} {dT} \Bigl (\frac {V} {U ^ {1/3}} \Bigr) =0 \,, \eqno(B.6)
$$
$$
-\frac {d} {dr} \Bigl [(1-kr^2) ^ {1/2} r^2\Bigr] +2 (1-kr^2) ^ {-1/2} r=0 . \eqno(B.7)
$$
For components $ \nu =2 $ and $ \nu =3 $ Eqs.~$(B.4)$ are fulfilled
identically.
From Eqs.~$ (B.6) $ and $ (B.7) $ it follows
$$
V/U ^ {1/3} = \mbox {const} = \beta^4\ne 0, \quad k=0 .
\eqno(B.8)
$$
Thus, the RTG \textbf {uniquely leads us to the flat spatial (Euclidean) geometry}
of the Universe.

Supposing
$$
a^2=U ^ {1/3},
\eqno(B.9)
$$
we obtain
$$
ds^2 = \beta^6\Bigl [d\tau_g^2-\Bigl (\frac {\, a \,} {\beta} \Bigr) ^2 (dr^2+r^2d\theta^2+r^2\sin^2\theta \,
d\phi^2) \Bigr] . \eqno(B.10)
$$
Here the following quantity
$$
d\tau_g =\Bigl (\frac {\, a \,} {\beta} \Bigr) ^3dT \eqno(B.11)
$$
determines the rate of slowing down of a rate of time at the presence
of a gravitational field in comparison to inertial time $T $.
The common constant numerical factor $ \beta^6 $ in an interval
$ds^2$ equally increases both time, and space variables.
It does not reflect dynamics of development of the Universe,
but determines the time in the Universe and its spatial scale.
The time in the Universe is determined by value $d\tau$ as  a
time-like part of  interval $ds^2$
$$
d\tau = \beta^3d\tau_g=a^3dT,
\eqno(B.12)
$$
$$
ds^2 =d\tau^2-\beta^4a^2 (\tau) (dr^2+r^2d\theta^2+r^2\sin^2\theta \, d\phi^2)
.
\eqno(B.13)
$$

The energy-momentum tensor of matter in the effective Riemannian space is as follows
$$
T _ {\mu\nu} = (\rho +p) U_\mu U_\nu -g _ {\mu\nu} p \,,
\eqno(B.14)
$$
where $ \rho $ and $p $ are correspondingly the density and the pressure of matter at the rest system of it, and $U_\mu $ is its velocity. As for
interval $(B.13)$
$g_{0i}$ and $R_{0i}$ are equal to zero it
follows from Eq.~$ (B.3) $ that
$$
T _ {0i} =0 \; \;\mbox {and} \; \; U_i=0 .
\eqno(B.15)
$$
It means that at the inertial system, defined by interval $(B.2)$, the matter
stays at rest during the evolution of Universe. The immovability of matter
in the homogeneous and isotropic Universe (distracting from pecular
velocities of galaxies) in some way corresponds to the early (before Fridman)
ideas by A.\, Einstein on the Universe.

So-called ``expansion of the Universe'', observed through the redshift, is caused not by a motion of matter, but by changing in
due rate of the gravitational field.
This note should be meant, when the accepted terminology ``expansion of the Universe'' is used.

Describing interval $ (B.13) $ in proper time $ \tau $ the interval
of initial Minkowski space $(B.2)$ will accept the following form
$$
d\sigma^2 =\frac {1} {a^6} d\tau^2-dr^2-r^2 (d\theta^2 + \sin^2\theta \, d\phi^2) . \eqno(B.16)
$$
On the base of Eqs.~$(B.13)$ and $(B.16)$, after taking into account that
$$
R _ {00} = -3\frac {\ddot {\, a \,}} {a}, \quad R _ {11} = \beta^4 (a\ddot {a} +2\dot {a} ^2) \,, \eqno(B.17)
$$
$$
T _ {00} -\frac {\, 1 \,} {2} g _ {00} T =\frac {\, 1 \,} {2} (\rho +3p), \quad T _ {11} -\frac {\, 1 \,} {2} g _
{11} T =\frac {\, 1 \,} {2} \beta^4 a^2 (\rho -p) \,, \eqno(B.18)
$$
from Eqs.~$(B.3)$ it is discovered
$$
\frac {\, 1 \,} {a} \frac {d^2 a} {d\tau^2} = -\frac {4\pi G} {3} \Bigl (\rho + \frac {3p} {c^2} \Bigr) -\frac {\,
1 \,} {6} (mc) ^2\Bigl (1-\frac {1} {a^6} \Bigr) \,, \eqno(B.19)
$$
$$\Bigl (\frac {\, 1 \,} {a} \frac {da} {d\tau} \Bigr) ^2
= \frac {8\pi G} {3} \rho (\tau) -\frac {1} {12} (mc) ^2 \Bigl (2-\frac {3} {a^2\beta^4} + \frac {1} {a^6} \Bigr).
\eqno(B.20)
$$
Differentiating $(B.20)$ over $\tau$ and using $(B.19)$, we obtain
$$
-\frac {\, 1 \,} {a} \frac {da} {d\tau} = \frac {1} {3\Bigl (\rho +\ds\frac {p} {c^2} \Bigr)} \frac {d\rho}
{d\tau} . \eqno(B.21)
$$
The above equation can also be immediately derived  from the covariant conservation
law: $$ \nabla_\mu (\sqrt {-g} \, T ^ {\mu\nu})
= \pa_\mu (\sqrt {-g} \, T ^ {\mu\nu}) + \varGamma ^\nu _ {\alpha\beta} (\sqrt
{-g} \, T ^ {\alpha\beta}) =0 \,,
\eqno(B.22)
$$
where $ \varGamma ^\nu _ {\alpha\beta} $ are Christoffel symbols of the effective
Riemannian space.

Let us write Eq.~$(B.21)$ in the following form
$$
-\frac {3p} {c^2} =a\frac {d\rho} {da} +3\rho . \eqno(B.23)
$$
Taking into account $(B.23)$ we can present $(B.19)$ as follows
$$
\frac {\, 1 \,} {a} \frac {d^2 a} {d\tau^2} = \frac {4\pi G} {3} \Bigl (a\frac {d\rho} {da} +2\rho\Bigr) -\frac
{\, 1 \,} {6} (mc) ^2\Bigl (1-\frac {1} {a^6} \Bigr) . \eqno(B.24)
$$
This equation can be written in the following form
$$
\frac {d^2a} {d\tau^2} = -\frac {dV} {da} \,, \eqno(B.25)
$$
where
$$
V = -\frac {4\pi G} {3} a^2\rho +\frac {(mc) ^2} {12} \Bigl (a^2 +\frac {1} {2a^4} \Bigr) . \eqno(B.26)
$$
Multiplying both parts of Eq.~$(B.25)$ onto $da/d\tau $, we
get
$$
\frac {d} {d\tau} \Bigl[\frac {\, 1 \,} {2} \Bigl (\frac {da} {d\tau} \Bigr) ^2+V\Bigr] =0 \,, \eqno(B.27)
$$
or
$$
\frac {\, 1 \,} {2} \Bigl (\frac {da} {d\tau} \Bigr) ^2+V=E =\mbox {const}. \eqno(B.28)
$$
Comparing $(B.20)$ and $(B.28)$, we obtain
$$
\beta^{\, 4} = (mc) ^2/8E .
\eqno(B.29)
$$
Thus, the constant $\beta^{\, 4}$ is determined by
integral of motion $E$. The expression $(B.28)$ reminds us the energy of a unit
mass. If the quantity $a$ had the dimensionality of length, the first term in $(B.28)$ would correspond to the kinetic energy, and second --- to the
potential one. From the causality  requirement (\ref {eq3}) it follows that
$a_{\max}=\beta$,  integral of motion $E$ is different from zero, but it is
very  small [8].

\end {document}